\documentclass[11pt,a4paper]{article}
\usepackage{styleArxiv}
\usepackage{amsmath}
\usepackage{amssymb}
\usepackage{eurosym}
\usepackage{graphicx}
\usepackage{color}
\usepackage{bm}
\begin{document}

%%%%%%%%%%%%%%%%%% title page information %%%%%%%%%%%%%%%%%%
\title{Three-dimensional Accelerating Electromagnetic Waves}
\author[a,1]{Miguel A. Bandres,} 
\note{http://www.mabandres.com/}
\author[b]{Miguel A. Alonso,}
\author[c]{Ido Kaminer}
\author[c]{and Mordechai Segev}

\affiliation[a]{Instituto Nacional de Astrof\'isica,~\'Optica y Electr\'onica,\\
Tonantzintla, Puebla 72840, Mexico}
\affiliation[b]{The Institute of Optics, University of Rochester,\\Rochester, NY 14627, USA}
\affiliation[c]{Physics Department and Solid State Institute, Technion,\\ Haifa 32000, Israel}

\emailAdd{bandres@gmail.com} %% email address is required

%\homepage{http://www.mabandres.com} %% author's URL, if desired


\abstract{
 We present a general theory of three-dimensional nonparaxial spatially-accelerating waves of the Maxwell equations. These waves constitute a two-dimensional structure exhibiting shape-invariant propagation along semicircular trajectories. We provide classification and characterization of possible shapes of such beams, expressed through the angular spectra of parabolic, oblate and prolate spheroidal fields.  
Our results facilitate the design of accelerating beams with novel structures, broadening scope and potential applications of accelerating beams.}
\maketitle 
%\ocis{(070.7345) Wave propagation; (070.3185) Invariant optical fields;
%(350.7420) Waves; (350.5500) Propagation; (260.2110) Electromagnetic optics.}

%%%%%%%%%%%%%%%%%%%%%%% References %%%%%%%%%%%%%%%%%%%%%%%%%

%%%%%%%%%%%%%%%%%%%%%%%%%%  body  %%%%%%%%%%%%%%%%%%%%%%%%%%
\section{Introduction}

The concept of self-accelerating beam, which was introduced into the domain of optics in 2007 \cite{Siviloglou,Dogariu}, has generated much follow-up and many new discoveries and applications. Generally, the term ``accelerating beams'' is  now used in conjunction with wave packets that preserve their shape while propagating along curved trajectories. The phenomenon arises from interference: 
the waves emitted from all points on the accelerating beam interfere in the exact manner that maintains a propagation-invariant structure, bending along a curved trajectory. This beautiful phenomenon requires no waveguiding structure or external potential, appearing even in free-space as a result of pure interference. The first optical accelerating beam, the paraxial Airy beam, was proposed and observed in 2007 \cite{Siviloglou,Dogariu}. Since then, research on accelerating beams has been growing rapidly, leading to many intriguing ideas and applications ranging from particle and cell micromanipulation \cite{Dholakia2}, light-induced  curved plasma channels \cite{Plasma}, self-accelerating nonlinear beams \cite{NLO}, self-bending electron beams \cite{ArieElectron} to accelerating plasmons \cite{Plasmon1} and applications in laser micromachining \cite{Dudley2}. Following the research on spatially-accelerating beams, similar concepts have been studied also in the temporal domain, where temporal pulses self-accelerate in a dispersive medium \cite{Siviloglou,a,b,c} up to some critical point determined by causality \cite{c}. Interestingly, shape-preserving accelerating beams were also found in the nonlinear domain \cite{d} in a variety of nonlinearities ranging from Kerr, saturable and quadratic media \cite{d,e,g,h} to nonlocal nonlinear media \cite{g}.

In two-dimensional (2D) paraxial systems (including the propagation direction and one direction transverse to it), the one-dimensional Airy beams are the only exactly shape-preserving solution to paraxial wave equation with accelerating properties. However, in three-dimensional (3D) paraxial systems, two separable solutions are possible: two-dimensional Airy beams \cite{Dogariu} and accelerating parabolic beams \cite{ApA,Apa2}. Furthermore, it has been shown \cite{AcB} that any function on the real line can be mapped to an accelerating beam with a different transverse shape. This allows creating paraxial accelerating beams with special properties such as reduced transverse width and beams with a transverse rainbow-like profile having a finite width, instead of a long tail, in the accelerating direction.

However, until 2012, the concept of accelerating beams was restricted to the paraxial regime, and the general mindset was that accelerating wave packets are special solutions for Schr\"{o}dinger-type equations, as they were originally conceived in 1979 \cite{BerryBalazs}.   
This means that the curved beam trajectory was believed to be restricted to small (paraxial) angles. In a similar vein, paraxiality implies that the transverse structure of paraxial accelerating beams cannot have small features, on the order of a few wavelengths or less. At the same time, reaching steep bending angles and having small scale features is fundamental in areas like nanophotonics and plasmonics, hence searching for shape-preserving accelerating nonparaxial wave packets was naturally expected.
Indeed, recent work \cite{Kaminer} has overcome the paraxial limit finding shape-preserving accelerating solutions of the Maxwell equations. These beams propagate along semi-circular trajectories \cite{Kaminer,i} that can reach, with an initial ``tilt'', almost $180\,^{\circ}$ turns \cite{KaminerOPN}. Subsequently, 2D nonparaxial accelerating wave packets with parabolic \cite{Weber,Zhang} and elliptical \cite{Aleahmad, Zhang} trajectories were found. Also, fully 3D nonparaxial accelerating beams were proposed, based on truncations or complex apodization of spherical, oblate and prolate spheroidal fields \cite{SW,Aleahmad}.
Finally, nonparaxial accelerating beams were suggested in nonlinear media \cite{Kaminer:12,j}.

All of this recently found plethora of nonparaxial accelerating beams suggest there might be a broader theory of self-accelerating beams of the three-dimensional Maxwell equations: a general formulation encompassing all the particular examples of \cite{AcB,Aleahmad}, and generalizing them to a unified representation. Such a theory could once and for all, answer several questions about the phenomenon of self-accelerating beams. For example, what kind of beam structures can display shape-preserving bending? What are the fundamental limits on their feature size and acceleration trajectories? %Are they always carrying infinite power, or can we find finite-power accelerating beams without the need for apodization (which always limits the range of shape-invariant propagation)? 
What trajectories would such beams follow?

Here, we present a theory describing the entire domain of 3D nonparaxial accelerating waves that propagate in a semicircle. These electromagnetic wave packets are monochromatic solutions to the Maxwell equations and they propagate in semicircular trajectories reaching asymptotically a $90\,^{\circ}$ bending in a quarter of a circle. We show that solutions exist with the polarization essentially perpendicular to their bending direction towards the path's center of curvature. In their scalar form, these waves are exact time-harmonic solutions of the wave equation. As such, they have implications to many linear wave systems in nature. We propose a classification and characterization of possible shapes of these accelerating waves, expressed through the angular spectra of parabolic, oblate and prolate spheroidal fields. We find novel transverse distributions, such as the nonparaxial counterpart of a 2D paraxial Airy beam, and accelerating beams that instead of a long tail, have a finite width (of a few wavelengths) in the transverse direction to the propagation direction that bends in a circle, among others.

\section{Three-dimensional nonparaxial accelerating waves}

We begin our analysis by considering the 3D Helmholtz
equation $\left( \partial _{xx}+\partial _{yy}+\partial _{zz}+k^{2}\right)
\psi =0$ where $k$ is the wave number. In free space, the solution of
the Helmholtz equation can be described in terms of plane waves through its
angular spectral function $A(\theta ,\phi )$ as 
\begin{equation}
\psi \left( \boldsymbol{r}\right) =\int A(\theta ,\phi )\exp \left( ik\boldsymbol{r}%
\cdot \boldsymbol{u}\right) \mathrm{d}\Omega ,  \label{spectra}
\end{equation}%
where $\boldsymbol{u}=\left( \sin \theta \sin \phi ,\cos \theta ,\sin \theta \cos \phi
\right) $ is a unit vector that runs over the unit sphere, and $\mathrm{d}%
\Omega =\sin \theta \mathrm{d}\theta \mathrm{d}\phi $ is the solid angle
measure on the sphere.

To search for wavepackets that are shape-preserving and whose
trajectory resides on a semicircle, it is convenient to start with solutions
whose trajectory resides on a full circle, i.e., solution with rotational
symmetry. These solutions have an intensity profile that is exactly
preserved over planes containing the $y$-axis and therefore they will have
defined angular momentum $J_{y}=-i\left( xd_{z}-zd_{x}\right)$ along this
axis. This operator acts on the spectral function as $J_{y}=-i\partial
_{\phi }$; hence, the spectral function of a rotationally symmetric solution
must satisfy $-i\partial _{\phi }A=mA$. In this way, any rotationally
symmetric wave must have a spectral function of the form $A(\theta ,\phi
)=g(\theta )\mathrm{exp}\left( im\phi \right) ,$ where $m$ is a positive
integer and $g\left( \theta \right) $ is any complex function in the
interval $[0,\pi ]$.

Although these rotationally symmetric fields are shape-invariant and travel in
a closed circle, they are composed of forward- (positive $k_{z}$, i.e., $\phi
\in \lbrack -\pi /2,\pi /2]$) and backward- (negative $k_{z}$, i.e., $\phi
\in \lbrack \pi /2,3\pi /2]$) propagating waves. Creating such rotationally-symmetric beams would require launching two pairs of counter-propagating beams (or two counter-propagating beams each with an initial tilt of virtually $90\,^{\circ}$ angle). Here, we are interested in beams that can be launched from a single plane. We therefore limit the integration in Eq.~(\ref{spectra}) to the forward semicircle $\phi \in \lbrack -\pi /2,\pi /2]$ resulting in a forward-propagating wave with accelerating characteristics that can be created by a standard optical system, i.e.,
\begin{figure}[t]
\centering\includegraphics[width=15cm]{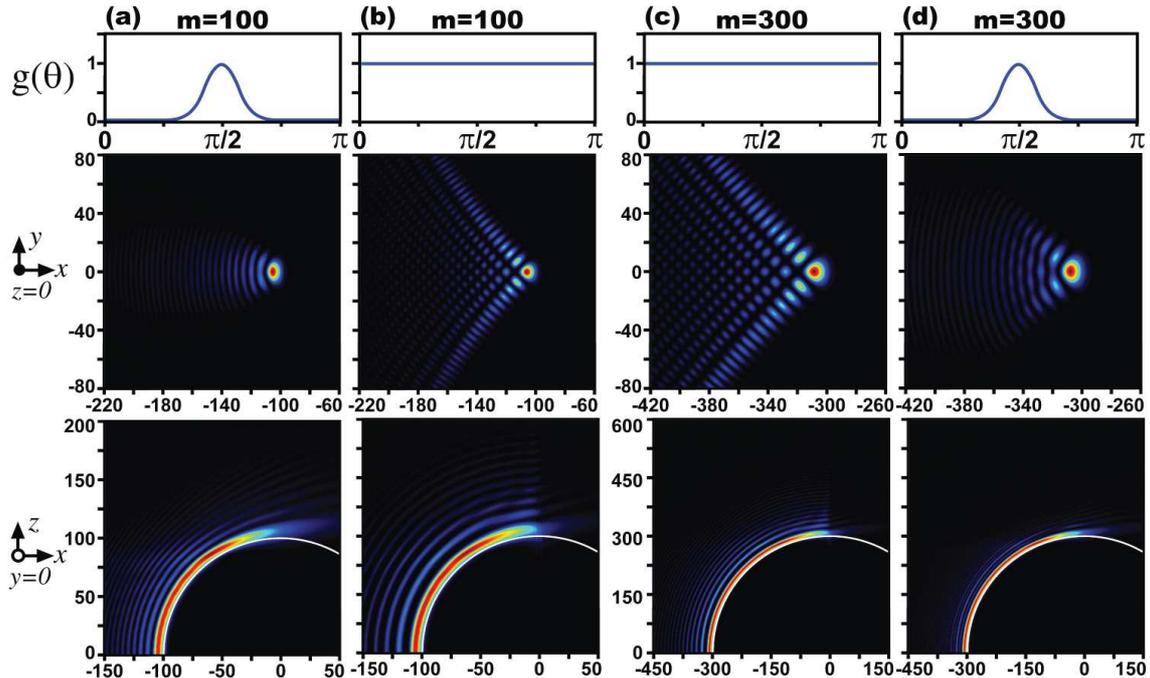}
\caption{Intensity cross-sections of three-dimensional nonparaxial accelerating beams and their corresponding generating functions $g(\theta)$.  
Top row: Amplitudes of the generating functions as a function of the $k$-space angle $\theta$. Middle row: Intensity cross-section at $z=0$ presenting the shape-invariant profile of each beam. Bottom row: Top-view plot showing the intensity cross-section at plane $y=0$ highlighting the circular trajectory. All lengths are in units of $k^{-1}$.}
\label{fig1}
\end{figure}

\begin{equation}
\psi\left( \boldsymbol{r}\right) =\int_{0}^{\pi }\sin \theta\mathrm{d}\theta\int_{-\pi /2}^{\pi
/2}\mathrm{d}\phi g(\theta )\mathrm{exp}\left( im\phi \right) \exp \left( ik\boldsymbol{r}\cdot 
\boldsymbol{u}\right)  ,  \label{Waves}
\end{equation}%
where now $m$ can be any positive real number (not necessarily an integer), because we are no longer restricted by periodic boundary conditions.
In this way, any function $g(\theta )$ generates a nonparaxial
accelerating wave with a different transverse distribution.
Furthermore, by construction, all these waves share the same
accelerating characteristics: their maxima propagate along a semicircular path of
radius slightly larger than $m/k$, while approximately preserving their 2D transverse shape up to almost $90\,^{\circ}$ bending angles. These characteristics are reminiscent of broken rotational
symmetry. Also, because larger angular momentum gives better spatial
separation of the counterpropagating parts of a rotational field, our
nonparaxial accelerating waves with larger $m$ are shape-invariant to larger propagation distances, but
their rate of bending will be slower, i.e., they follow a larger circle.

Figure \ref{fig1} shows several transverse-field distributions and
propagation of nonparaxial accelerating waves with their corresponding $%
g(\theta )$. As we can see in Fig. \ref{fig1} the semicircular propagation
path has a radius $m/k$. It is possible to double the angle of bending (from $90\,^{\circ}$ to $180\,^{\circ}$) by propagating these waves from $z<0$.
In this case the waves have a bending angle opposite to the direction of bending and depict full semicircles. Moreover, notice that the propagation
characteristics are independent of $g(\theta )$, and that $g(\theta )$ only
controls the shape of the transverse profile. As a consequence, on one hand, if we superpose accelerating waves with different values of $m$, they will
interfere during propagation, leading to families of periodic self-accelerating waves \cite{Kaminer,KaminerOPN}. On the other hand, waves with the same $m$ will propagate with the same propagation constant, hence they will maintain their relative phase as in the initial plane, and preserve their nondiffractive behavior.

Our construction of accelerating waves extends into the nonparaxial regime the
construction of paraxial accelerating beams in \cite{AcB}, where it is shown
that any function $\ell \left( k_{y}\right) $ on the real line can be mapped
to an accelerating beam. This is in direct analogy to our function $g\left( \theta \right)$ of the nonparaxial case. While in the paraxial case the bending (i.e., transverse acceleration) is controlled by an overall scale parameter, in the nonparaxial
case it is controlled by $m$ as described previously.

\begin{figure}[t]
\centering\includegraphics[width=14cm]{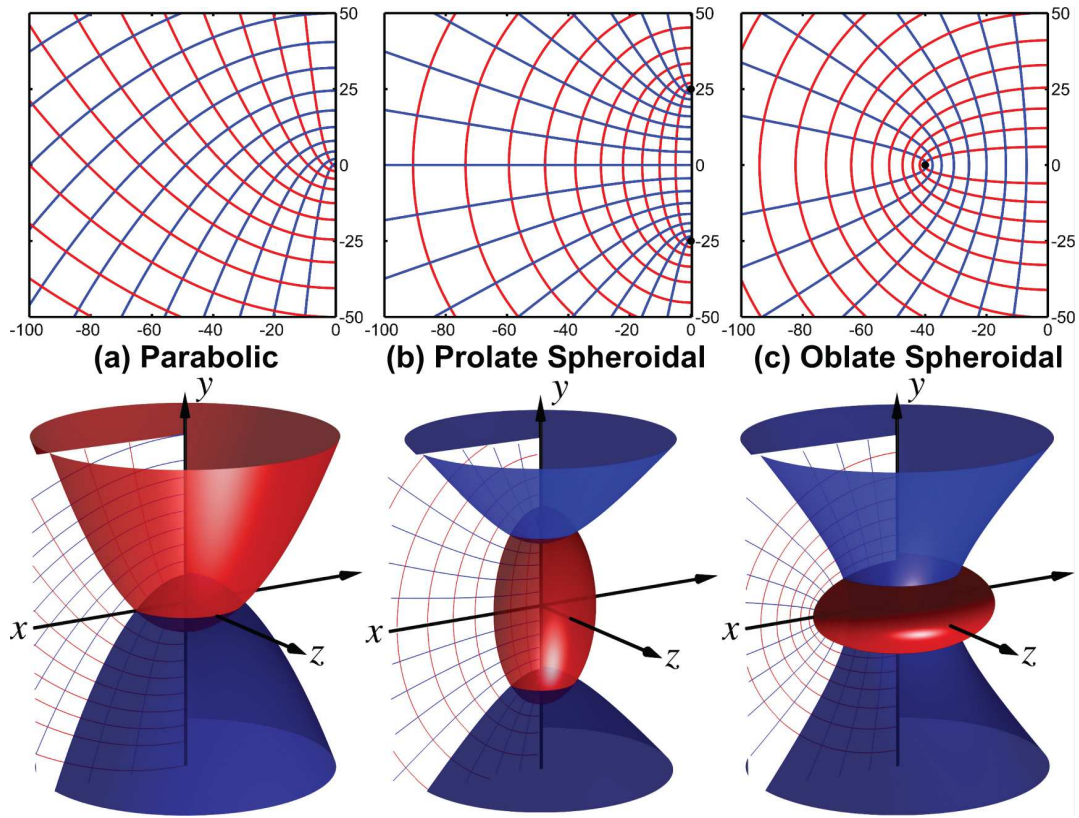}
\caption{Three-dimensional rotational coordinate systems of the Helmholtz equation.}
\label{FigCoordinates}
\end{figure}

Although any function $g(\theta )$ can generate an accelerating wave, it is
not straightforward to visualize (ab initio) the features of the transverse
profile that that function will generate. For this reason, we propose to use
the $g(\theta )$ functions associated with rotationally symmetric separable
solutions of the Helmholtz equation. As it is known from \cite{Miller}, there are
only four rotationally symmetric solutions to this equation, corresponding to
the spherical, parabolic, prolate spheroidal and oblate spheroidal
coordinate systems, depict in Fig. \ref{FigCoordinates}. The advantages of borrowing the spectral function of
these solutions is that we can create complete families of nonparaxial
accelerating waves and readily characterize their transverse structures.

The physical meaning of the separability of these solutions is that these
waves have three conserved physical constants. The first one is the
conservation of energy given by the Helmholtz equation, the second one is
the conservation of azimuthal angular momentum, and the third conserved
quantity is specific to each case and corresponds to generalization of
the total angular momentum for each coordinate. This last symmetry will
characterize the transverse profile of the waves, i.e., their caustics.
Interestingly, a family of rays sharing the same conserved constants have equivalent caustics to our accelerating waves.

The spectral functions used here correspond to fields that are separable solutions of the wave equation, expressible in terms of known special functions in the case when the plane wave superposition involves components traveling in all possible directions. However, here we are limiting the integration to forward propagating waves in order to describe fields that would be easy to generate with standard optical setups. It must be noted that this truncation does cause the resulting fields not to be expressible in closed form, although the solutions are shown to essentially preserve the field profiles of the separable solutions. An alternative approach that would allow preserving the closed-form expressions while suppressing backward propagating components is that of performing imaginary displacements on the separable solutions, as discussed in \cite{SW}.

In the next sections, we describe in detail the parabolic, prolate spheroidal
and oblate spheroidal nonparaxial accelerating waves. The spherical
accelerating waves have been presented in \cite{SW,Aleahmad} and nonparaxial
accelerating waves based on spatial truncations of the full prolate and
oblate spheroidal wave functions where presented in \cite{Aleahmad}. Although
for large values of $m$ our waves can be approximated by those of \cite{Aleahmad},
our Fourier space approach allows us to generate the waves without the need
of calculating neither the radial functions nor the coordinate system.

\section{Parabolic accelerating waves}

\begin{figure}[t]
\centering\includegraphics[width=15cm]{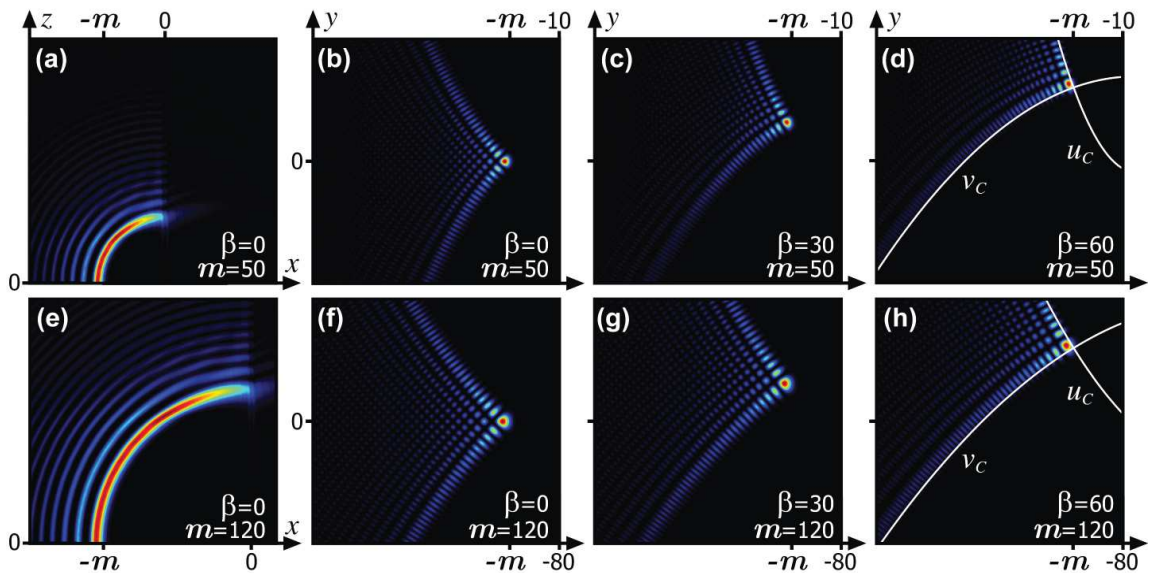}
\caption{Parabolic accelerating waves with different ``translation'' values of $\beta$. (a,e) Intensity cross-section at the $y=0$ plane, (b-d,f-h) intensity profiles at $z=0$ planes. The white line parabolas in (d,h) depict the caustic cross sections. All sections are of size $200\times 200$ and all lengths are in units of $k^{-1}$.}
\label{FigParabolic}
\end{figure}

We generate the parabolic accelerating waves by evaluating Eq.~(\ref{Waves})
with the following spectral function 
\begin{equation}
g_{\beta }\left( \theta \right) =\frac{1}{2\pi }\frac{\left[ \tan \left(
\theta /2\right) \right] ^{i\beta}}{\sin \theta },\qquad -\infty <\beta
<\infty ,
\end{equation}
where $\beta$ is a continuous ``translation'' parameter of the waves.
The transverse field distributions at $z=0$ of the parabolic accelerating
waves are shown in Fig. \ref{FigParabolic}. As one can see, these profiles
resemble the ones of the 2D paraxial Airy beams \cite{Siviloglou,Dogariu}; this is
because the parabolic coordinate system looks like a Cartesian coordinate
system rotated $45^{\circ}$ near a coordinate patch at $y\approx 0,$ $%
\left\vert x\right\vert \gg 1$, as shown in Fig. \ref{FigCoordinates}(a).
The main lobe of the waves is located near $x=-m/k,$ $y=\beta /k$.  The
fundamental mode is $\beta=0$ and as $\beta$ increases the waves ``translate'' in the $y$-axis.
This is consistent with the result of \cite{AcB} where it is shown that the paraxial
2D Airy beams are orthogonal under translations perpendicular to the
direction of acceleration.
Notice that in our case for $\beta \neq 0$ there is also a ``tilt'' in the caustic accompanied by a change in the spacing of the fringes along the caustic sheets. Note that this ``tilt'' does not change the direction of propagation, thus the acceleration is still horizontal in Fig.~\ref{FigParabolic}, and not in the direction to which the intensity pattern points, as it might seem at first. This resembles a paraxial 2D Airy beam with different scale parameters for each of the constituent Airy functions. 
 As shown in Figs. \ref{FigParabolic}(a) and \ref%
{FigParabolic}(e), the parabolic accelerating waves present a single
intensity main lobe that follows a circular path of radius slightly larger
than $m/k$.

By separation of variables the Helmholtz equation can be broken into ordinary
differential equations \cite{Miller} with an effective potential for each
coordinate. The turning point of these effective potentials will give the
caustics of the solutions. We find that our accelerating waves share
these caustics in a form reminiscent of the broken symmetries. In this way, we find that the caustics of the parabolic accelerating waves are given by 
\begin{equation}
u_{C}^{2}=\left( -\beta +\sqrt{\beta ^{2}+m^{2}}\right) /k,\quad
v_{C}^{2}=\left( \beta +\sqrt{\beta ^{2}+m^{2}}\right) /k,
\end{equation}%
where the parabolic coordinates $\left[ u,v,\phi \right] ,$ are defined as 
\begin{equation}
x=uv\sin \phi ,\quad y=\frac{1}{2}\left( u^{2}-v^{2}\right) ,\quad z=uv\cos
\phi ,
\end{equation}%
where $u\in \lbrack 0,\infty ],v\in \lbrack 0,\infty ],$ $\phi \in \lbrack
0,2\pi ).$ The caustic cross sections are depicted in Fig. \ref%
{FigParabolic}(d) and \ref{FigParabolic}(h); by rotating these around the $y$%
-axis one gets the caustic surfaces which are two paraboloids, see Fig. \ref%
{FigCoordinates}(a).

\section{Prolate spheroidal accelerating waves}
\begin{figure}[t]
\centering\includegraphics[width=15cm]{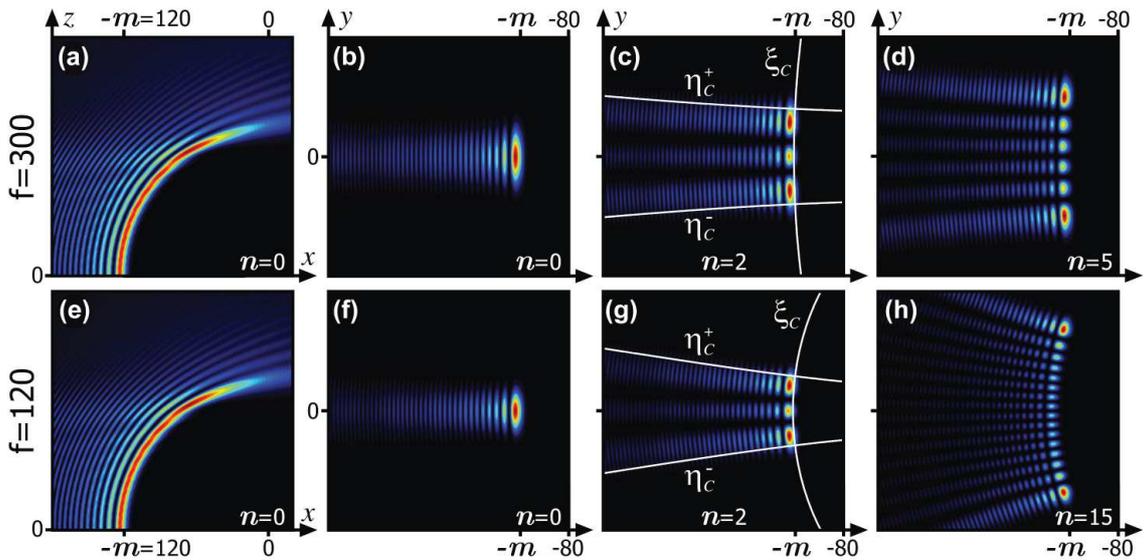}
\caption{Prolate spheroidal accelerating beams of different orders $n$. (a,e) Intensity cross-section at the $y=0$ plane. (b-d,f-h) Intensity profiles at the $z=0$ plane. The beam of order $n$ has exactly $n+1$ stripes. The white line hyperbolas and ellipses in (c,g) depict the caustic cross sections. All subfigures are for $m=120$, of size $200\times 200$, and all lengths are in units of $k^{-1}$.}
\label{FigProlate}
\end{figure}

We construct the prolate spheroidal accelerating waves by evaluating Eq.~(\ref%
{Waves}) with the following spectral function%
\begin{equation}
g_{n}^{m}\left( \theta ;\gamma \right) =S_{m+n}^{m}\left( \cos \theta
,\gamma \right) ,\qquad \gamma \equiv kf,
\end{equation}%
where the foci of the prolate spheroidal coordinate system are at $(0,\pm f,0)$, $m=0,1,2,\ldots ,$ $n=0,1,2,\ldots ,$ and $S_{l}^{m}\left( \bullet
\right) $ is the spheroidal wave function \cite{Spheroidal} that satisfies%
\begin{equation}
\frac{d}{d\nu }\left[ \left( 1-\nu ^{2}\right) \frac{d}{d\nu }%
S_{l}^{m}\left( \nu ,\gamma \right) \right] +\left( \Lambda _{l}^{m}-\gamma
^{2}\nu -\frac{m^{2}}{1-\nu ^{2}}\right) S_{l}^{m}\left( \nu ,\gamma \right)
=0,
\end{equation}%
where $\Lambda _{l}^{m}\left( \gamma \right) $ is the eigenvalue of the
equation.

Several transverse intensity distributions at $y=0$ and $z=0$ of the prolate spheroidal accelerating waves are shown in Fig. \ref{FigProlate}. The waves have a
definite parity with respect to the $y$-axis, which is given by the parity of $n$%
. The order $n$ of the waves corresponds to the number of hyperbolic nodal
lines at the $z=0$ plane, and the width of the waves in the $y$-axis increases as $n$
increases. As shown in Figs. \ref{FigProlate}(a) and \ref{FigProlate}(e),
the prolate accelerating waves have two main lobes (or a single lobe for $n=0$%
) that follow a circular path of radius slightly larger than $m/k,$ i.e.,
the degree $m$ of the waves controls their propagation characteristics.

To understand the behavior of the prolate waves for different $f$, let us analyze how the prolate spheroidal coordinate system behaves
as a function of $f.$ As $f\rightarrow 0$ the foci coalesce and the prolate spheroidal
coordinates tend to the spherical coordinates, while in the other extreme, as 
$f\rightarrow \infty $ the prolate spheroidal coordinates tend to the
circular cylindrical ones. 
Irrespective of the value of $f$, we find that the entire beam is always restricted to $\sqrt{x^2+z^2}>m/k$, This limit can be understood as a centrifugal force barrier. 

Using this notation, we divide the prolate accelerating beams into three regimes:
\begin{itemize}
\item For $m\gtrsim kf,$ the prolate accelerating waves resemble the
spherical accelerating waves described in \cite{SW,Aleahmad}, cf. Figs. \ref%
{FigProlate}(f,h) and Figs. 2(j,l) of \cite{SW}.

\item For $m<kf$, the waves are located in a coordinate patch that
approximates a Cartesian system, hence the prolate accelerating waves
take the form $A(x)H(y)$, where $A(x)$ is an accelerating function and $H(y)$ is a
function that retains its form upon propagation and has finite extend.

\item For $m\ll kf$ , the prolate spheroidal coordinates tend to the
circular cylindrical ones, and the prolate accelerating waves tend to
the product of a ``half-Bessel'' wave \cite{Kaminer} in the $x$-coordinate
times a sine or cosine in the $y$-coordinate.
\end{itemize}

To complete the characterization of the prolate accelerating beams, we find the caustic surfaces to be a prolate spheroid and two-sheet hyperboloids given by%
\begin{align}
\sin ^{2}\eta _{C}^{+}&=\frac{-\left( \Lambda -\gamma ^{2}\right) +\sqrt{%
\left( \Lambda -\gamma ^{2}\right) ^{2}+4\gamma ^{2}m^{2}}}{2\gamma ^{2}}, 
\\
\sinh ^{2}\xi _{C}&=\frac{\left( \Lambda -\gamma ^{2}\right) +\sqrt{%
\left( \Lambda -\gamma ^{2}\right) ^{2}+4\gamma ^{2}m^{2}}}{2\gamma ^{2}},
\end{align}%
and $\eta _{C}^{-} =\pi -\eta _{C}^{+}$, where the prolate spheroidal coordinates $\left[ \xi ,\eta ,\phi \right] ,$
are defined as 
\begin{equation}
x=f\sinh \xi \sin \eta \sin \phi ,\quad y=f\cosh \xi \cos \eta ,\quad
z=f\sinh \xi \sin \eta \cos \phi ,
\end{equation}%
and $\xi \in \lbrack 0,\infty ],\eta \in \lbrack 0,\pi ],$ $\phi \in \lbrack
0,2\pi ).$ The caustic cross sections are depicted in Fig. \ref{FigProlate}%
(c) and \ref{FigProlate}(g); by rotating this around the $y$-axis one gets
the caustic surfaces, see Fig. \ref{FigCoordinates}(b).

\section{Oblate spheroidal accelerating waves}
\begin{figure}[tbp]
\centering\includegraphics[width=15cm]{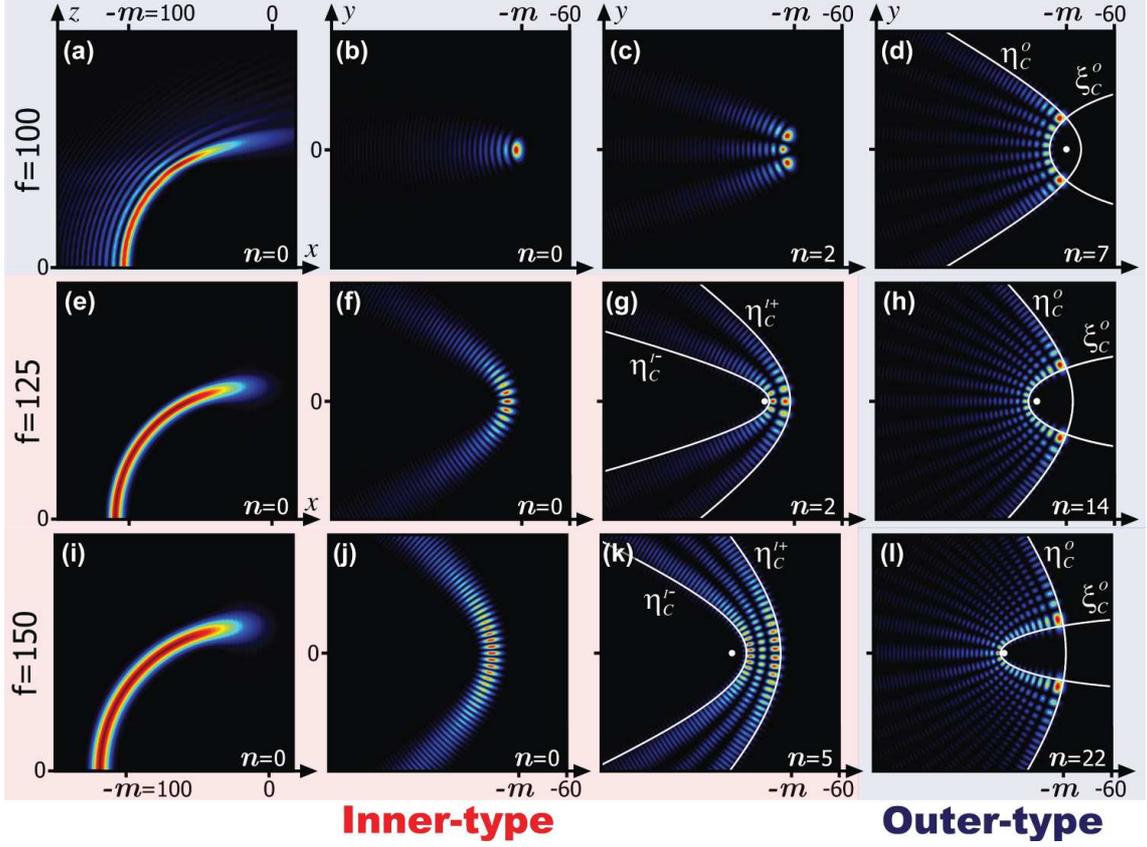}
\caption{Oblate spheroidal accelerating beams of outer-type and inner-type. (a,e,i) Intensity cross-section at the $y=0$ plane. (b-d,f-h,j-l) Intensity profiles at the $z=0$ plane. The white line hyperbolas and ellipses in (g,h,k,l) depict the caustic cross sections. The white dots correspond to the foci. All subfigures are for $m=100$, of size $200\times 200$, and all lengths are in units of $k^{-1}$.} 
\label{FigOblate}
\end{figure}

The oblate spheroidal accelerating waves are given by evaluating Eq.~(\ref%
{Waves}) with the following spectral function%
\begin{equation}
g_{n}^{m}\left( \theta ;i\gamma \right) =S_{m+n}^{m}\left( \cos \theta
,i\gamma \right) ,\qquad \gamma \equiv kf,
\end{equation}%
where $f$ is the radius of the focal ring in the $y=0$ plane, $m=0,1,2,\ldots ,$ and $n=0,1,2,\ldots .$ Notice that the prolate and
oblate spectral functions are related by the transformation $\gamma
^{2}\rightarrow -\gamma ^{2}$, 
yet the two families exhibit different physical properties, that resemble each other only in the spherical limit ($m\gg kf$). 

By studying the caustics of the oblate accelerating waves we find that they have
two types of behavior according to the value of the eigenvalue of $%
S_{m+n}^{m}\left( \cos \theta ,i\gamma \right)$, $\Lambda _{m+n}^{m}\left(
i\gamma \right) $. On the one hand, if $\Lambda _{m+n}^{m}\left( i\gamma \right) > m^{2},$ the caustic is composed of an oblate spheroid and a hyperboloid of revolution; we will
call these waves outer-type. On the other hand, if $\Lambda _{m+n}^{m}\left( i\gamma \right)
< m^{2},$ the caustic is composed of two hyperboloids of revolution; we will call these
waves inner-type [see Fig.~\ref{FigOblate}~(f-g,j-k)]. Interestingly, in general this last condition is only fulfilled if $kf>m$. Because $\Lambda_{m+n}^{m}\left( i\gamma \right)$ increases as $n$ increases, for any $kf>m$ there is a maximum value of $n$ for inner-type waves and for higher $n$ values the waves become outer-type. This transition from inner-type to outer-type as $n$ increases is depicted in middle and bottom rows of Fig.~\ref{FigOblate}.

\subsection{Outer-type}
Outer-type oblate accelerating waves are depicted in Fig.~\ref{FigOblate}. The degree $m$ of the waves controls their propagation characteristics because their two main lobes (or single lobe for $n=0$) follows a circular path of radius slightly larger than $m/k$ [see Fig.~\ref{FigOblate})(a)]. The order $n$ gives its parity with respect to the $y$-axis and corresponds to the number of hyperbolic nodal lines at the $z=0$ plane. One of the two cusps that $n>0$ oblate waves have, can be suppressed by combining three of these field as in \cite{SW}, i.e., $\Psi^m_n-i/2\left(\Psi^m_{n+1}-\Psi^m_{n-1}\right) \label{UP}$. Notice that $n=0$ outer-type waves are very thin (several wavelenghts), even more confined in the $y$-axis than the parabolic and prolate accelerating waves, cf. Fig.~\ref{FigOblate}(b) and Fig.~\ref{FigParabolic}(b), Fig.~\ref{FigProlate}(b); this gives these type of waves a potential advantage in applications.

Near a coordinate patch at $\mid x \mid \approx f$ and $y\approx 0$ the transverse coordinates look like a parabolic system, see Fig.~\ref{FigCoordinates}(c). Then for $m = f $ the oblate accelerating waves become the nonparaxial version of the paraxial accelerating parabolic beams in \cite{ApA,Apa2}, cf. Fig.~\ref{FigOblate}(a,c,e) and Fig.1(b,c,d) of \cite{ApA}.

The caustics of the outer-type oblate accelerating waves are given by%
\begin{align}
\sin ^{2}\eta _{C}^{O}&=\frac{\left( \Lambda +\gamma^{2}\right) -\sqrt{\left(
\Lambda -\gamma^{2}\right) ^{2}+4\gamma^{2}m^{2}}}{2\gamma^{2}}, \\
\cosh ^{2}\xi
_{C}^{O}&=\frac{\left( \Lambda +\gamma^{2}\right) +\sqrt{\left( \Lambda
+\gamma^{2}\right) ^{2}-4\gamma^{2}m^{2}}}{2\gamma^{2}},
\end{align}%
where the oblate spheroidal coordinates $\left[ \xi ,\eta ,\phi \right] ,$
are defined as 
\begin{equation}
x=f\cosh \xi \cos \eta \sin \phi ,\quad y=f\sinh \xi \sin \eta ,\quad
z=f\cosh \xi \cos \eta \cos \phi ,
\end{equation}%
and $\xi \in \lbrack 0,\infty ],\eta \in \lbrack -\pi /2,\pi /2],$ $\phi \in
\lbrack 0,2\pi ).$ The caustic cross sections are depicted in Fig. \ref{FigOblate}(d), \ref{FigOblate}(h), and \ref{FigOblate}(l); by rotating this
around the $y$-axis one gets the caustic surfaces, which are an oblate
spheroid and a hyperboloid of revolution, see Fig. \ref{FigCoordinates}(c).

\subsection{Inner-type}
Inner-type oblate accelerating waves form $\lceil (n+1)/2\rceil$ hyperpolic stripes that separate two regions of darkness [see Fig.~\ref{FigOblate}(f,g,j,k)] and therefore their topological structure is different than all the other waves presented in this work. First, the caustic of these waves does not present a cusp. Also, the intensity cross section at the $y=0$ plane of the $n=0$ inner-type wave only presents a single lobe of several wavelengths width, instead of a long tail of lobes present in all the other accelerating beams, cf. Fig.~\ref{FigOblate}(e,i) and Fig.~ \ref{FigOblate}(a). Moreover, the position of the maximum is no longer near $x=-m$ but at some $x<-m$. The maximum amplitude remains constant during propagation until it decays very close to $90^\circ$ of bending; this behavior is completely different than other accelerating waves that present a small oscillation of their maximum during propagation - compare Fig.~\ref{FigOblate}(e,i) and Fig.~\ref{FigOblate}(a). Finally, these waves have definite parity with respect to the $y$-axis, which is given by the parity of $n$. For example, the waves with $n=2$ [see Fig.~\ref{FigOblate}(g)] and $n=3$ both form two parabolic stripes, but have opposite parity. If we combine these waves of opposite parity, i.e., $\psi_n \pm i \psi_{n+1}$, where $n$ is even, we can create continuous stripes of light that will also carry momentum along the hyperbolic stripes at a given $z$-plane.

The caustics of inner-type oblate accelerating waves are given by%
\begin{align}
\sin ^{2}\eta _{C}^{I+}&=\frac{\left( \Lambda +\gamma ^{2}\right) -\sqrt{%
\left( \Lambda -\gamma ^{2}\right) ^{2}+4\gamma ^{2}m^{2}}}{2\gamma ^{2}}%
, \\
\sin ^{2}\eta _{C}^{I-}&=\frac{\left( \Lambda +\gamma ^{2}\right) +%
\sqrt{\left( \Lambda +\gamma ^{2}\right) ^{2}-4\gamma ^{2}m^{2}}}{2\gamma
^{2}}.
\end{align}%
The caustics cross sections are depicted in Fig. \ref{FigOblate}(g) and \ref%
{FigOblate}(k); by rotating this around the $y$-axis one gets the caustic
surfaces which are two hyperboloids of revolution.

\section{Vector solutions}
While up to this point our work has dealt with scalar waves, full vector accelerating waves can be readily constructed from these results by using the Hertz vector potential formalism. This formalism shows that an electromagnetic field in free-space can be defined in terms of a single auxiliary vector potential \cite{Stratton}. In this way, if the auxiliary Hertz vector potentials $\boldsymbol{\Pi}_{e,m}$ satisfy the vector Helmholtz equations, i.e., $\nabla ^{2}\boldsymbol{\Pi }_{e,m}+k^{2}\boldsymbol{\Pi }_{e,m}=0$, one can recover the electromagnetic field components by
\begin{equation}
\mathbf{H}=i\omega \epsilon \nabla \times \boldsymbol{\Pi }_{e},\qquad \mathbf{E}%
=k^{2}\boldsymbol{\Pi }_{e}+\nabla \left( \nabla \cdot \boldsymbol{\Pi }_{e}\right), \label{E-type}
\end{equation}
which are called electric type waves or  
\begin{equation}
\mathbf{E}=-i\omega \mu \nabla \times \boldsymbol{\Pi }_{m},\qquad \mathbf{H%
}=k^{2}\boldsymbol{\Pi }_{m}+\nabla \left( \nabla \cdot \boldsymbol{\Pi }_{m}\right), \label{M-type}
\end{equation}%
which are called magnetic type waves. Therefore, we can create electromagnetic accelerating waves with different vector polarizations by setting $\boldsymbol{\Pi}_{e,m}=\psi \widehat{v}$, where $\psi $ is any of our scalar accelerating waves and $\widehat{v}$
is any unit vector of a Cartesian coordinate system.

\begin{figure}[t]
\centering\includegraphics[width=15cm]{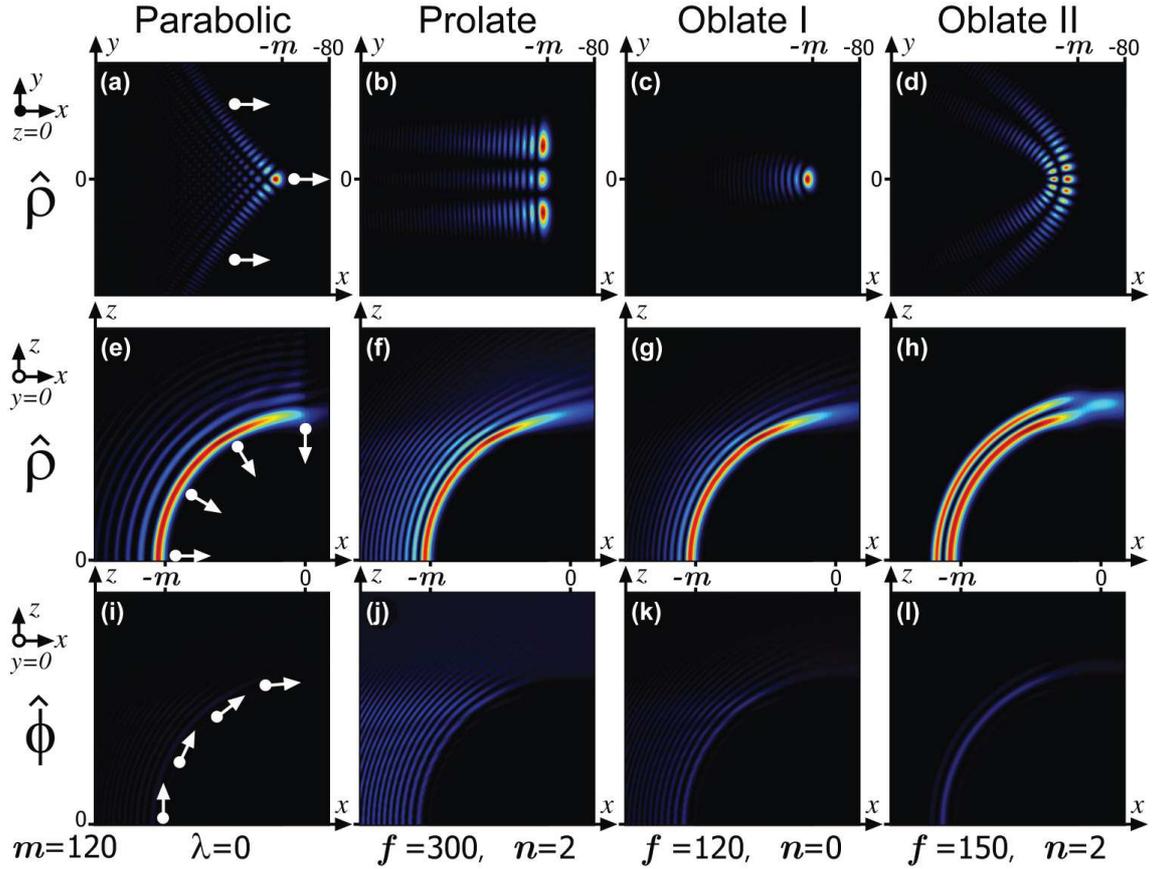}
\caption{
Comparison between vector parabolic, prolate, and oblate accelerating beams. Each row shows the electric field intensity of the radial $\widehat{\rho }$ and angular $\widehat{\phi }$ components of a single electromagnetic accelerating wave at the $y=0$ and $z=0$ planes, over sections of size $200\times 200$. The white arrows in the first row depict the polarization. All lengths are in units of $k^{-1}$.}
\label{FigVector}
\end{figure}

The Hertz vector potential $\boldsymbol{\Pi }_{m}=\psi \widehat{y}$ is of
special interest because it gives electromagnetic accelerating waves that
share the same characteristics as our scalar accelerating waves. The
electric field given by $\boldsymbol{\Pi }_{m}=\psi \widehat{y}$ is 
\[
\mathbf{E=}i\omega \mu \left( -\widehat{\rho }\frac{\partial _{\phi
}}{\rho }+\widehat{\phi }\partial _{\rho }\right) \psi ,
\]%
where $\left( \rho ,\phi ,y\right) $ are circular cylindrical coordinates
related to our Cartesian coordinates by $\left( x,y,z\right) =\left( \rho
\sin \phi ,y,\rho \cos \phi \right) $. This cylindrical coordinate system
for the polarization is useful because it shows that the radial component
is dominant: As we already showed, in the region of interest our scalar
waves behave approximately as $\psi \sim F\left( \rho /k,y/k\right)
e^{im\phi }$. Then $\rho ^{-1}\partial _{\phi }\psi \sim im\rho ^{-1}\psi $,
and because the maximum of $\psi $ is around $\rho \sim m/k$ the
amplitude of the maximum of the radial component is approximately $k\psi $%
. Now, $\partial _{\rho }\psi \sim \left( F^{\prime }/F\right) k\psi $ and $%
F^{\prime }/F\ll 1$ around the main lobes of $\psi $. 
This allows us to show that the radial component $\widehat{\rho }$ is the dominant one. This behavior was confirmed by comparing both components numerically. Hence, the polarization of the accelerating beams is perpendicular to the direction of propagation that bends in a circle. Physically, this
makes sense, since the polarization must be perpendicular to the propagation direction of each plane-wave constituent of the beam, and in the case of our accelerating electromagnetic waves the radial polarization is always perpendicular to the
direction of propagation of the whole wave packet that bends in a circle. Figure~\ref{FigVector} shows the radial and angular components of the electric field of
several accelerating electromagnetic waves at the $x=0$ and $z=0$ planes;
notice that the radial component preserves the shape and
propagation characteristics of the scalar accelerating waves.

\section{Conclusion}

To summarize, we presented a general theory of three-dimensional nonparaxial accelerating electromagnetic waves, displaying a large variety of transverse distributions. These waves propagate along a semicircular trajectory while maintaining an invariant shape. In their scalar form, these waves are exact time-harmonic solutions of the wave equation; therefore they have implications to many linear wave systems in nature such as sound, elastic and electron waves. Moreover, in their electromagnetic form, these families of waves span the full vector solutions of the Maxwell equations, in several different representations, each family presenting a different basis for this span. By using the angular spectrum of parabolic, oblate and prolate spheroidal fields, we gave a classification and characterization of the possible transverse shape distributions of these waves. As a final point, because our accelerating waves are nonparaxial, they can bend to steep angles and have features of the order of the wavelength; characteristics that are necessary and desirable in areas like nanophotonics, plasmonics, and micro-particle manipulation.

\section{Acknowledgments}
MAA acknowledges support from the National Science Foundation (PHY-1068325). 
MAB acknowledge useful correspondence with W. Miller Jr.


\begin{thebibliography}{10}
\newcommand{\enquote}[1]{``#1''}

\bibitem{Siviloglou}
G.~A. Siviloglou and D.~N. Christodoulides, \enquote{Accelerating finite energy
  {A}iry beams,} Opt. Lett. \textbf{32}, 979--981 (2007).

\bibitem{Dogariu}
G.~A. Siviloglou, J.~Broky, A.~Dogariu, and D.~N. Christodoulides,
  \enquote{Observation of accelerating {A}iry beams,} Phys. Rev. Lett.
  \textbf{99}, 213901 (2007).

\bibitem{Dholakia2}
J.~Baumgartl, G.~M. Hannappel, D.~J. Stevenson, D.~Day, M.~Gu, and K.~Dholakia,
  \enquote{Optical redistribution of microparticles and cells between
  microwells,} Lab Chip \textbf{9}, 1334--1336 (2009).

\bibitem{Plasma}
P.~Polynkin, M.~Kolesik, J.~V. Moloney, G.~A. Siviloglou, and D.~N.
  Christodoulides, \enquote{Curved plasma channel generation using ultraintense
  {Airy} beams,} Science \textbf{324}, 229--232 (2009).

\bibitem{NLO}
T.~Ellenbogen, N.~{Voloch-Bloch}, A.~{Ganany-Padowicz}, and A.~Arie,
  \enquote{Nonlinear generation and manipulation of {Airy} beams,} Nature
  Photonics \textbf{3}, 395--398 (2009).

\bibitem{ArieElectron}
N.~Voloch-Bloch, Y.~Lereah, Y.~Lilach, A.~Gover, and A.~Arie,
  \enquote{Generation of electron {Airy} beams,} Nature \textbf{494}, 331--335
  (2013).

\bibitem{Plasmon1}
A.~Minovich, A.~Klein, N.~Janunts, T.~Pertsch, D.~Neshev, and Y.~Kivshar,
  \enquote{Generation and {Near-Field} imaging of {Airy} surface plasmons,}
  Physical Review Letters \textbf{107}, 116802 (2011).

\bibitem{Dudley2}
A.~Mathis, F.~Courvoisier, L.~Froehly, L.~Furfaro, M.~Jacquot, P.~A. Lacourt,
  and J.~M. Dudley, \enquote{Micromachining along a curve: Femtosecond laser
  micromachining of curved profiles in diamond and silicon using accelerating
  beams,} Applied Physics Letters \textbf{101}, 071110 (2012).

\bibitem{a}
A.~Chong, W.~H. Renninger, D.~N. Christodoulides, and F.~W. Wise,
  \enquote{Airy--{B}essel wave packets as versatile linear light bullets,}
  Nature photonics \textbf{4}, 103--106 (2010).

\bibitem{b}
D.~Abdollahpour, S.~Suntsov, D.~G. Papazoglou, and S.~Tzortzakis,
  \enquote{Spatiotemporal {A}iry light bullets in the linear and nonlinear
  regimes,} Phys. Rev. Lett. \textbf{105}, 253901 (2010).

\bibitem{c}
I.~Kaminer, Y.~Lumer, M.~Segev, and D.~N. Christodoulides, \enquote{Causality
  effects on accelerating light pulses,} Opt. Express \textbf{19}, 23132--23139
  (2011).

\bibitem{d}
I.~Kaminer, M.~Segev, and D.~N. Christodoulides, \enquote{Self-accelerating
  self-trapped optical beams,} Phys. Rev. Lett. \textbf{106}, 213903 (2011).

\bibitem{e}
I.~Dolev, I.~Kaminer, A.~Shapira, M.~Segev, and A.~Arie, \enquote{Experimental
  observation of self-accelerating beams in quadratic nonlinear media,} Phys.
  Rev. Lett. \textbf{108}, 113903 (2012).

\bibitem{g}
Y.~Hu, Z.~Sun, D.~Bongiovanni, D.~Song, C.~Lou, J.~Xu, Z.~Chen, and
  R.~Morandotti, \enquote{Reshaping the trajectory and spectrum of nonlinear
  {A}iry beams,} Opt. Lett. \textbf{37}, 3201--3203 (2012).

\bibitem{h}
R.~Bekenstein and M.~Segev, \enquote{Self-accelerating optical beams in highly
  nonlocal nonlinear media,} Opt. Express \textbf{19}, 23706--23715 (2011).

\bibitem{ApA}
M.~A. Bandres, \enquote{Accelerating parabolic beams,} Opt. Lett. \textbf{33},
  1678--1680 (2008).

\bibitem{Apa2}
J.~A. Davis, M.~J. Mintry, M.~A. Bandres, and D.~M. Cottrell,
  \enquote{Observation of accelerating parabolic beams,} Opt. Express
  \textbf{16}, 12866--12871 (2008).

\bibitem{AcB}
M.~A. Bandres, \enquote{Accelerating beams,} Opt. Lett. \textbf{34}, 3791--3793
  (2009).

\bibitem{BerryBalazs}
M.~V. Berry and N.~L. Balazs, \enquote{Nonspreading wave packets,} Am. J. Phys.
  \textbf{47}, 264--267 (1979).

\bibitem{Kaminer}
I.~Kaminer, R.~Bekenstein, J.~Nemirovsky, and M.~Segev, \enquote{Nondiffracting
  accelerating wave packets of {M}axwell's equations,} Phys. Rev. Lett.
  \textbf{108}, 163901 (2012).

\bibitem{i}
F.~Courvoisier, A.~Mathis, L.~Froehly, R.~Giust, L.~Furfaro, P.~A. Lacourt,
  M.~Jacquot, and J.~M. Dudley, \enquote{Sending femtosecond pulses in circles:
  highly nonparaxial accelerating beams,} Opt. Lett. \textbf{37}, 1736--1738
  (2012).

\bibitem{KaminerOPN}
I.~Kaminer, E.~Greenfield, R.~Bekenstein, J.~Nemirovsky, M.~Segev, A.~Mathis,
  L.~Froehly, F.~Courvoisier, and J.~M. Dudley, \enquote{Accelerating beyond
  the horizon,} Opt. Photon. News \textbf{23}, 26--26 (2012).

\bibitem{Weber}
M.~A. Bandres and B.~M. Rodr\'iguez-Lara, \enquote{Nondiffracting accelerating
  waves: Weber waves and parabolic momentum,} New Journal of Physics
  \textbf{15}, 013054 (2013).

\bibitem{Zhang}
P.~Zhang, Y.~Hu, T.~Li, D.~Cannan, X.~Yin, R.~Morandotti, Z.~Chen, and
  X.~Zhang, \enquote{Nonparaxial {M}athieu and {W}eber accelerating beams,}
  Phys. Rev. Lett. \textbf{109}, 193901 (2012).

\bibitem{Aleahmad}
P.~Aleahmad, M.-A. Miri, M.~S. Mills, I.~Kaminer, M.~Segev, and D.~N.
  Christodoulides, \enquote{Fully vectorial accelerating diffraction-free
  {H}elmholtz beams,} Phys. Rev. Lett. \textbf{109}, 203902 (2012).

\bibitem{SW}
M.~A. Alonso and M.~A. Bandres, \enquote{Spherical fields as nonparaxial
  accelerating waves,} Opt. Lett. \textbf{37}, 5175--5177 (2012).

\bibitem{Kaminer:12}
I.~Kaminer, J.~Nemirovsky, and M.~Segev, \enquote{Self-accelerating
  self-trapped nonlinear beams of {M}axwell's equations,} Opt. Express
  \textbf{20}, 18827--18835 (2012).

\bibitem{j}
P.~Zhang, Y.~Hu, D.~Cannan, A.~Salandrino, T.~Li, R.~Morandotti, X.~Zhang, and
  Z.~Chen, \enquote{Generation of linear and nonlinear nonparaxial accelerating
  beams,} Opt. Lett. \textbf{37}, 2820--2822 (2012).

\bibitem{Miller}
C.~P. Boyer, E.~G. Kalnins, and W.~Miller~Jr, \enquote{Symmetry and separation
  of variables for the {H}elmholtz and {L}aplace equations,} Nagoya Math. J
  \textbf{60}, 35–80 (1976).

\bibitem{Spheroidal}
L.-W. Li, M.-S. Leong, T.-S. Yeo, P.-S. Kooi, and K.-Y. Tan,
  \enquote{Computations of spheroidal harmonics with complex arguments: A
  review with an algorithm,} Phys. Rev. E \textbf{58}, 6792--6806 (1998).

\bibitem{Stratton}
J.~Stratton, \emph{Electromagnetic theory}, vol.~33 (Wiley-IEEE Press, 2007).

\end{thebibliography}
\end{document}